\documentstyle[aps,prl,multicol,epsfig]{revtex}

\newcommand{\be}{\begin{eqnarray}}
\newcommand{\ee}{\end{eqnarray}}
\def\ket#1{|#1\rangle}

\draft
\tighten

\begin{document}
\title{Laser probing of atomic Cooper pairs}

\author{P.\ T\"orm\"a$^{1,2,3}$ and P.\ Zoller$^1$}
\address{$^1$Institute for Theoretical Physics, University of Innsbruck,
Technikerstra{\ss}e 25, A-6020 Innsbruck, Austria\\
$^2$Helsinki Institute of Physics, P.O.Box 9, 
FIN-00014 University of Helsinki, Finland\\
$^3$Laboratory of Computational Engineering, P.O.Box 9400, FIN-02015
Helsinki University of Technology, Finland}
\maketitle

\begin{abstract} 
We consider a gas of attractively interacting cold Fermionic atoms
which are manipulated by laser light. The laser induces a transition
from an internal state with large negative scattering length to one
with almost no interactions. The process can be viewed as a tunneling of
atomic population between the superconducting and the normal states of the
gas. It can be used to detect the BCS-ground state and to measure the
superconducting order parameter.
\end{abstract}

\pacs{05.30.Fk, 32.80.-t, 74.25.Gz}

\begin{multicols}{2}[]

Successful cooling of trapped Fermionic $^{40}$K atoms down to
temperatures where the Fermi degeneracy sets in was reported recently
\cite{Jin99,Janne99}. This breakthrough opens up new opportunities for
studying fundamental quantum statistical and many-body physics. A
major advantage of Fermionic atoms compared to electrons in condensed
matter is the richness of their internal energy structure and the
possibility to accurately and efficiently manipulate these energy
states by laser light and magnetic fields. Furthermore, atomic gases
are dilute and weakly interacting thus offering the ideal tool for
developing and experimentally testing theories of many-body quantum physics.
A major goal is to observe the predicted \cite{Stoof,You99} BSC-transition
for Fermionic atoms -- this would compare to the experimental
realization of atomic Bose-Einstein condensates \cite{Bose}.  It is
still, however, an open question how to observe the BCS-transition,
because the value of the superconducting order parameter (gap) is
expected to be small and electro-magnetic phenomena such as the
Meissner effect do not take place.

We propose a method to detect the existence of the BCS-ground state
and to measure the gap using laser light. The main difference in our
method compared to the proposals of using off-resonant light
scattering \cite{Weig99,Zhang99} is that the light is resonant, that
is, population is transferred from one internal state to
another. Furthermore, the final state of this process is chosen to
have negligible scattering length compared to the initial one: this
effectively creates a superconducting -- normal state interface across
which the atomic population can move. There is a conceptual analogy to
electron tunneling from a superconducting metal to a normal one,
although the physical systems and their describtions differ in an
essential way. Also non-optical phenomena, such as collective and
single particle excitations, have been proposed to be used for
observing the BCS-transition \cite{Baranov98,Baranov99,Bruun99}. The
specific feature of our method with respect to other suggested probes,
both optical and mechanical (modulating the trap frequency), is the
utilization of the superconducting -- normal state interface. The
advantage is that the normal state population can be initally zero,
thus even a small change in it is a significant relative
effect. Furthermore, our work offers a basic starting point for
describing any phenomenon related to the superconducting--normal
interface. For instance, our method can be
understood as an outcoupler. With small modifications, one
could also describe outcoupling of pairs instead of single 
atoms which would create an atomic beam with BCS-type statistics.

\narrowtext
\vbox{
\begin{figure}
\begin{center}
\epsfig{file=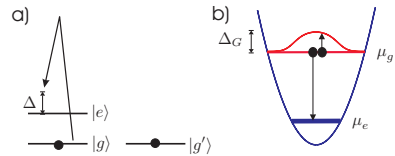,width=2.0in,angle=0} 
\end{center}
\caption{Probing of the gap in a gas of attractively interacting cold
Fermionic atoms. a) Laser excitation with the coupling $\Omega$ and
the detuning $\Delta$ transfers a Cooper paired atom from
the internal state $\ket{g}$ to the state $\ket{e}$.  b) The other
atom in the initial Cooper pair becomes an excitation in the BCS
state, therefore the laser has to provide also the additional gap
energy $\Delta_G$.  In this picture the Fermi levels $\mu_g$ and 
$\mu_e$ for the
internal states have been chosen to be different from each other but
they could also equal.} 
\end{figure} } 

We consider atoms with three internal states available, say $\ket{e}$,
$\ket{g}$, and $\ket{g'}$. They are chosen so that the interaction
between atoms in states $\ket{g}$ and $\ket{g'}$ is relatively strong
and attractive, and all other interactions are negligible. The laser
frequency is chosen to transfer population between $\ket{e}$ and
$\ket{g}$, but is not in resonance with any transition which could
move population away from the state $\ket{g'}$. If there is more or
less equal amount of atoms in states $\ket{g}$ and $\ket{g'}$ they can
form the superconducting BCS ground state; atoms in $\ket{e}$ are in
the normal state. For small intensities, the laser-interaction can be
treated as a perturbation, the unperturbed states being the normal and
the superconducting state. The transfer of atoms from $\ket{g}$ to
$\ket{e}$ is then analogous to tunneling of electrons from a normal
metal to a superconductor induced by an external voltage, which can be
used as a method to probe the gap and the density of states of the
superconductor \cite{Mahan}. In our case the tunneling is between
two internal states rather than two spatial regions, this
resembles the idea of internal Josephson-oscillations in
two-component Bose-Einstein condensates \cite{Williams99}.
Figure 1 illustrates the basic idea.  
The observable carrying essential information about the superconducting
state is the change in the population of the state $\ket{e}$, we call
this the current $I$.  Below we calculate the current first in the
homogeneous case and then assuming that the atoms are confined in a
harmonic trap.  Finally will discuss possibilities for 
experimental realization of the idea, and choosing the states
$\ket{e}$, $\ket{g}$, and $\ket{g'}$ for $^6$Li.

We define a two-component Fermion field
$\psi ({\bf x}) = 
(\psi_e({\bf x}) \quad \psi_g({\bf x}))^T$,
where $\psi_e$ and $\psi_g$ fulfil standard Fermionic
commutation relations. The fields $\psi_{e/g}$ can be expanded
using some basis functions (e.g. plane waves or trap wave functions) and 
corresponding creation and annihilation operators:
$
\psi_{e/g}({\bf x})= \sum_j c_j^{e/g} \phi_j^{e/g} 
({\bf x})
$.
The annihilation and creation operators fulfil $\{ {c_i^e}^\dagger,
c_j^g\}=0$ and $\{ c_i^{e/g\dagger},c_j^{e/g}\}=\delta_{ij}$. 
The two components of the field, corresponding to the internal
states $\ket{e}$ and $\ket{g}$, are coupled by a laser.
This can be either a direct excitation or a Raman process;
we denote the atomic energy level difference 
by $\omega_a$ ($\hbar\equiv1$), the laser frequency 
$\omega_L$ and the wave vector
$k_L$ -- in the case of a Raman process these are effective
quantities. In the rotating wave approximation the Hamiltonian reads
\be
H &=& H_e + H_{gg'} + \frac{\Delta}{2} \int d^3x \psi^\dagger ({\bf x}) \sigma_z
\psi ({\bf x}) \nonumber \\
&&+  \int d^3x \psi^\dagger ({\bf x})
\left( \begin{array}{cc}
0 & \Omega({\bf x}) \\
\Omega^*({\bf x}) & 0 
\end{array} \right) 
\psi ({\bf x}) .
\ee
Here $\Delta = \omega_a - \omega_L$ is the (effective) detuning, and
$\Omega({\bf x})$ contains the spatial dependence of the laser field
multiplied with the (effective) Rabi frequency.  The parts $H_e$ and
$H_{gg'}$ contain terms which depend only on $\psi_e$ or $\psi_g$, $\psi_{g'}$,
respectively. Possible spatial inhomogeneity,
e.g.\ from the trap potential, is also included in $H_e$ and
$H_{gg'}$.  

The observable carrying essential information about
the superconducting state is the rate of change in 
population of the $\ket{e}$ state. 
We may call it, after the electron tunneling analogy, 
the current $I(t) = - \langle \dot{N_e} \rangle$, where
$
N_e = \int d^3x  \psi_e^\dagger ({\bf x}) \psi_e ({\bf x})
$.
The current $I(t)$ is calculated considering the tunneling part of
the Hamiltonian, $H_T = H - (H_e + H_{gg'} + \Delta/2 (N_e - N_g))$ 
as a perturbation;
the current $I$ becomes the first order response 
to the external perturbation caused by the laser. We calculate
it both in the homogeneous case and in the case of harmonic confinement.
The calculations are done in the grand canonical ensemble, therefore 
the chemical potentials $\mu_g$ and $\mu_e$ are introduced. 
Also the detuning $\Delta$ acts like
a difference in chemical potentials, thus it becomes useful to 
define an effective quantity of the form $\tilde{\Delta} = \mu_e
- \mu_g + \Delta \equiv \Delta \mu + \Delta$. In the derivation
we assume finite temperature, but here we only quote the results
for $T=0$.

{\it Homogeneous case:} The assumption of spatial homogeneity is
appropriate when the atoms are confined in a trap
potential which changes very little compared to characteristic
quantities of the system, such as the coherence length and 
the size of the Cooper pairs.
In the homogeneous case we expand the Fermion fields $\psi_{e/g}$
into plane waves. The Hamiltonian becomes
\be
H &=& H_e + H_{gg'} + \frac{\Delta}{2} \sum_k [{c_k^e}^\dagger c_k^e
- {c_k^g}^\dagger c_k^g] \nonumber \\
&&+ \sum_{kl} [T_{kl} {c_k^e}^\dagger c_l^g
+ h.c.] ,  \label{ham2}
\ee
where
$
T_{kl} = \frac{1}{V} \int d^3x 
\Omega ({\bf x}) e^{i{\bf k} \cdot {\bf x}} e^{-i{\bf l} \cdot {\bf x}}
$.
We calculate the current $I=-\langle \dot{N_e} \rangle = -i \langle 
[H,N_e] \rangle$ treating $H_T$ as a perturbation: terms of higher
order than $H_T^2$ are neglected. Because we are interested in the
current between the superconducting and normal states, correlations
of the form $\langle c_e^\dagger c_e^\dagger c_g c_g \rangle$ 
(and $h.c.$) are omitted since they correspond to tunneling of 
pairs (Josephson current). The current can be written
\be
I = \int_{-\infty}^\infty dt \theta (t) 
(e^{-i\tilde{\Delta}t} \langle [A^\dagger(0),A(t)] \rangle
-e^{i\tilde{\Delta}t} \langle
[A(0),A^\dagger(t)] \rangle ) \nonumber
\ee
where $A(t) = \sum_{kl} T_{kl} {c_k^g}^\dagger(t) c_l^e(t)$, and
$c_l^{e/g}(t) = e^{iKt} c_l^{e/g} e^{-iKt}$ where $K=H-\mu_e N_e
-\mu_g N_g$. The two terms in the above equation have the form of
retarded and advanced Green's functions. These are evaluated
using Matsubara Green's functions techniques, which leads to 
$
I = \sum_{kl} |T_{kl}|^2 \int_{-\infty}^\infty \frac{d\epsilon}{2\pi}
[n_F(\epsilon) - n_F(\epsilon + \tilde{\Delta})]
A_g(k,\epsilon + \tilde{\Delta})A_e(l,\epsilon) , 
$
where $n_F$ are the Fermi distribution functions and $A_{g/e}$ 
are the spectral functions for the superconducting and normal
states. We use the standard expressions $A_{e}(l, \epsilon) = 
2\pi \delta (\epsilon - \xi_l)$ and $A_{g}(k, \epsilon +
\tilde{\Delta}) = 2 \pi [u_k^2 \delta (\epsilon + \tilde{\Delta}
- \omega_k) + v_k^2 \delta (\epsilon + \tilde{\Delta} + \omega_k)]$
\cite{Mahan}. Here $\xi_l = E_l - \mu_e$ and $u_k$, $v_k$ and 
$\omega_k$ are given by the Bogoliubov transformation.
Because we are interested in the tunneling out of 
the superconductor, only the term proportional to
$v_k^2$ is considered. The laser field is chosen to be a running wave, that
is $\Omega ({\bf x}) = \Omega e^{i {\bf k_L} \cdot {\bf x}}$. The
term $|T_{kl}|^2$ now produces a delta-function enforcing momentum
conservation. Note that this is very different from the 
assumption of a constant transfer matrix ($\sum_{kl} |T_{kl}|^2 
\longrightarrow |T|^2 \sum_{kl}$) made in the standard calculation
for tunneling of electrons over a superconductor -- normal metal
surface \cite{Mahan}. The final result becomes (assuming $T=0$)
\be
I = - 2\pi \Omega^2 \rho \theta (-\tilde{\Delta}-\omega_{\tilde{k}-k_L} -
\Delta \mu ) \frac{\omega_{\tilde{k}-k_L} -
\xi_{\tilde{k}-k_L}}{\omega_{\tilde{k}-k_L} 
+ \xi_{\tilde{k}-k_L}\left[1-\frac{k_L}{\tilde{k}}\right]}   \nonumber
\ee
where $\tilde{k}$ is given by the following energy conservation
condition:
$-\tilde{\Delta}+\omega_{\tilde{k}-k_L} + \xi_{\tilde{k}-k_L} = 0$,
$\omega_k = \sqrt{\xi_k^2 + \Delta_G^2}$, and $\rho$ is the
density of states which appears when the summation over momenta is
changed into an integration over energies.

The laser momentum $k_L$ can be very small compared to the momentum
of the atoms, especially in the case of a Raman process. By setting
$k_L=0$ the result becomes (choosing $\tilde{\Delta}<0$ i.e.\ current
from $\ket{g}$ to $\ket{e}$) 
\be
I = - 2 \pi \Omega^2 \rho \theta (\tilde{\Delta}^2 - \Delta_G^2 + 2
\Delta \mu \tilde{\Delta}) \frac{\Delta_G^2}{\tilde{\Delta}^2} .
\label{peak}
\ee

To understand the results in terms of physics, let us first consider
the case of equal chemical potentials $\Delta \mu = 0$:
\be
I =  - 2 \pi \Omega^2 \rho \theta (-\Delta - \Delta_G)
\frac{\Delta_G^2}{\Delta^2} .  \label{curr2}
\ee
In order to transfer one atom from the state $\ket{g}$ to $\ket{e}$
the laser has to break a Cooper pair. The minimum energy required for this
is the gap energy $\Delta_G$, therefore the current does not flow 
before the laser detuning provides this energy -- this is expressed
by the step function in (\ref{curr2}). As $|\Delta|$ increases
further, the current will decrease quadratically. This is because 
the case $|\Delta|=\Delta_G$ corresponds to the transfer of particles
with $p=p_F$, whereas larger $|\Delta|$ means larger momenta, and there
are simply fewer Cooper-pairs away from the Fermi surface. This
behaviour is very different from the electron tunneling where the current grows
as $\sqrt{(eV)^2 - \Delta_G^2}$ \cite{Mahan} (the voltage $eV$ corresponds to
the detuning $\Delta$ in our case) because all momentum states are coupled
to each other. 

When the chemical potentials are not equal the situation is more
complicated, but the basic features are the same: i) treshold for the
onset of the current given basically by the gap energy, and ii) further decay
of the current because the density of the states that can fulfil energy
and momentum conservation decreases.

{\it Harmonic confinement:}  
The trap wave functions provide the natural basis of expansion
when the atoms are confined in a harmonic trap. 
We will not, however, do this expansion from the beginning
of the calculations
but rather derive the current and the corresponding
(position dependent) Green's functions directly for the
Fermion fields $\psi_{e/g}$. The current becomes $I = 2 Im
[X_{ret}(-\tilde{\Delta})]$,
\be
&&X_{ret}(-\tilde{\Delta}) = \int_{-\infty}^\infty
\frac{d\epsilon}{2\pi} \int d^3 x \int d^3 x' \Omega^*({\bf x})
\Omega({\bf x'}) \nonumber \\ && [\tilde{A}_e({\bf x}, {\bf x'}, \epsilon) 
G_{adv}^g({\bf x'}, {\bf x}, \epsilon + \tilde{\Delta}) \nonumber \\ && 
+ G_{ret}^e({\bf x}, {\bf x'}, \epsilon - \tilde{\Delta})
\tilde{A}_g({\bf x'}, {\bf x}, \epsilon)] ,  \label{curr3}
\ee
and $\tilde{A}_{e/g}$ are defined
$\tilde{A}_{e/g}({\bf x}, {\bf x'}, \epsilon ) = i (G_{ret}^{e/g}({\bf
x}, {\bf x'}, \epsilon)
- G_{adv}^{e/g}({\bf x}, {\bf x'}, \epsilon))$.
In order to take the imaginary part of the expression (\ref{curr3})
we use the following Green's functions
\be
G_{adv}^{g}({\bf x'}, {\bf x}, \epsilon) &=& \sum_n \frac{u_n({\bf
x'})u_n^*({\bf x})}{\epsilon - \omega_n - i \delta}
+ \frac{v_n^*({\bf
x'})v_n({\bf x})}{\epsilon + \omega_n - i \delta}  \\
G_{ret}^{e}({\bf x}, {\bf x'}, \epsilon) &=& \sum_n \frac{\phi_n ({\bf x})
\phi_n^* ({\bf x'})}{\epsilon - \xi_n + i \delta} .
\ee
Here $u_n({\bf x})$, $v_n({\bf x})$ and $\omega_n$ are given by the 
Bogoliubov--DeGennes equations \cite{deGennes}.
We also use the fact that the trap wave functions $\phi_m({\bf x})$ are real. 
The derivation gives
\be
I &=& - 2\pi \sum_{n,m} \left|\int d^3 x \Omega ({\bf x}) u_n ({\bf x})
\phi_m({\bf x})\right|^2 [n_F(\omega_n) - n_F(\xi_m)] \nonumber \\ &&\delta (\xi_m +
\tilde{\Delta} - \omega_n) + \left|\int d^3 x \Omega ({\bf x}) v_n^* ({\bf x})
\phi_m({\bf x})\right|^2 \nonumber \\ && [n_F(-\omega_n) - n_F(\xi_m)] \delta (\xi_m +
\tilde{\Delta} + \omega_n)  .
\ee
Again, we consider only the term proportional to $v_n$, assume
zero temperature and use the known form for the trap energy $\xi_m
= m \Omega_t - \mu_e$ to obtain the final result
\be
I = \frac{2\pi}{\Omega_t} \sum_n \theta(-\tilde{\Delta} + \omega_n)
\left| \int d^3 x \Omega ({\bf x}) v_n^*
({\bf x}) \phi_M ({\bf
x})\right|^2  , \nonumber
\ee
where the quantum number $M=(\mu_e - \tilde{\Delta} + \omega_n)/\Omega_t$.
Although the result is not as transparent as in the homogeneous case,
the characteristic treshold behaviour is expressed by the step
function. Since the functions inside the absolute square are typically
oscillating ones, this term would restrict the values of $n$ in the
summation. We have checked that with strong approximations one obtains
again the homogeneous case result.

{\it Experimental realization:} Typical experimental parameters in the
case of $^6$Li are $\Omega_t/2\pi \sim 150Hz$ for the trap frequency
and $\mu_g \sim 100kHz$ for the chemical potential \cite{Stoof97}.
The order of magnitude of the gap is $\Delta_G/\epsilon_F \sim 0.1$
(see e.g.\ \cite{Weig99}). In Figure 2 we show the dependence of the
current on the detuning in the homogeneous case, for $\Delta_G = 0.1
\mu_g$. The sharp peak given by the result of Eq.(\ref{peak}) will be
broadened in an experimental situation. One of the main causes
of broadening is that it is difficult to fix the particle number
with high accuracy. We have simulated this by assuming that 
the particle number ($\mu_g$) varies from experiment to experiment
according to a Gaussian distribution (the width of the Gaussian
is used as a measure of the fluctuations), and averaged over the
different results. In these calculations we have introduced 
corresponding Gaussian fluctuations in the gap energy, because it is 
determined by the particle number. 
For as large as 10\% fluctuations  in the gap the peak is still
clearly visible although broadened and slightly shifted.

\narrowtext
\vbox{
\begin{figure}
\begin{center}
\epsfig{file=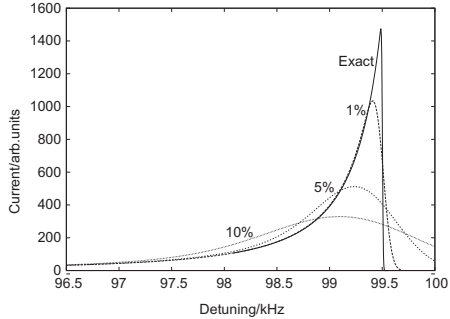,width=2.3in,angle=0}
\end{center} 
\caption{The absolute value of the
current $I=-\langle \dot{N}_e \rangle$ as a function of
the laser detuning $\Delta$. The chemical potentials are $\mu_g = 100
kHz$ and $\mu_e=0$. The current is calculated assuming 1\%, 5\%, and 10\%
fluctuations in the gap energy. Corresponding fluctuations
in the particle number ($\mu_g$) are taken into account using the
exponential dependence between the Fermi energy and the gap
energy. The unit of the current is arbitrary.  In the case of a normal
state, there would be a sharp peak at $\Delta = 100 = \mu_g -
\mu_e$. Due to the superconducting gap this peak is shifted and its
shape modified.}  
\end{figure}
} 

It is possible to prepare the desired initial number of
atoms in states $\ket{e}$, $\ket{g}$ and $\ket{g'}$ very accurately by
applying suitable incoherent laser pulses --- this is one of the
advantages of atomic Fermi gases compared to condensed matter systems.
The number of atoms in the states $\ket{g}$ and $\ket{g'}$ should be
nearly equal to produce the superconducting state, whereas the number
of atoms in state $\ket{e}$ is not critical. In the simplest case
the state $\ket{e}$ would be empty in the beginning. The use of
our method then requires to know $\mu_g$ rather accurately, since
it appears in the tunable argument $\tilde{\Delta} = \Delta \mu
+ \Delta = -\mu_g + \Delta$ of the current $I(\tilde{\Delta})$.
If this is difficult, it might be better to use an incoherent
pulse to prepare initially $\Delta \mu = \mu_e - \mu_g = 0$.
In this case the chemical potential dependence disappears
from the expressions in the homogeneous case. 
 
Finally, we discuss how to choose the states $\ket{e}$, $\ket{g}$ and
$\ket{g'}$ for a real atom (for instance $^6$Li).  It will probably be
difficult to find states in which the atoms do not interact at all. A
more likely solution is to choose the states so that the interaction
between all of them is weak and repulsive, except between $\ket{g}$
and $\ket{g'}$ strong and attractive. Even a weak and attractive
interaction instead of weak and repulsive would do, because the gap,
critical temperature, and other essential BCS quantities have an
exponential dependence on the scattering length, so the effect becomes
easily negligible.  Particularly $^6$Li seems to offer a variety of
interaction strengths between different states.  As shown in
\cite{Houbiers98}, in a magnetic field of about $B\simeq 0.02 T$ the
scattering lenght of several low field seeking hyperfine states is as
strong as $a_S \sim - 2000 a_0$ whereas the scattering length between
two high field seeking states is around zero. This is an example of
the potential of modifying the scattering lengths by magnetic
fields. In this case the trap should be optical in order to confine
also the high field seekers. Alternatively, in a magnetic trap atoms
in these states (corresponds to the non-interacting state $\ket{e}$ in
our calculation) would simply fly out --- this could even make the
detection of the current easier.  The detection of the atoms in the
state $\ket{e}$, that is, the measurement of our basic observable $I$,
could be done for instance by direct resonance fluorescence, or when a
Raman process is used, indirectly from the intensity difference of the
two Raman beams.

In conclusion, we have proposed a method to observe the
superconducting order parameter (gap) in a gas of cold Fermionic
atoms. The idea is based on creating a normal phase -- superconductor
interface by coupling internal states of different scattering lengths
by a laser.  The advantage of our scheme is the sensitivity to even
small currents of atomic population caused by the laser
interaction. Furthermore, our results extend beyond measuring the
order parameter: they can be used as a starting point in describing
various processes related to the transfer of atoms between two gas
components (in superconducting or normal states).
If both
of the internal states used were strongly interacting, one could
perhaps observe Josepson tunneling of pairs from one superconductor to
another. With small modifications, our results also describe the
functioning of an outcoupler from a superconducting gas of Fermionic
atoms.

{\it Acknowledgements} We thank H.T.C.\ Stoof for useful discussions.
PT acknowledges the support by the TMR programme of the
European Commission (ERBFMBICT983061) and the Academy of Finland
(projects 42588 and 47140).

\end{multicols}
\end{document}